# Further Evidence for an Increase of the Mean Mass of the Highest-Energy Cosmic-Rays with Energy

A A Watson
School of Physics and Astronomy
University of Leeds
Leeds LS2 9JT, UK

a.a.watson@leeds.ac.uk

**Abstract:** Sokolsky and D'Avignon have recently reported an examination of a variety of measurements that relate to the question as to how the mass composition of the highest-energy cosmic-rays evolves with energy. They assert that cosmic rays arriving from the Northern Hemisphere have a different mass composition from those arriving from the Southern Hemisphere, implying a diversity of sources of high-energy cosmic-rays in the two hemispheres. Were this conclusion to be correct, it would have profound implications for theories of cosmic-ray origin and would influence planning of future projects. Their claim thus merits careful scrutiny. In this paper their analysis is examined in detail with the verdict being that evidence for a North/South difference is not proven, a conclusion supported by other data from the Northern and Southern Hemispheres. However, what is of major importance is that the study of Sokolsky and D'Avignon provides long-awaited confirmation of the claim that the mean mass of cosmic rays increases with energy above ~3 EeV made by the Pierre Auger Collaboration in 2014.

## 1. Introduction

In a recent paper Soklosky and D'Avignon (2021 – hereafter SDA) have examined a variety of measurements that yield information on the evolution of the mass composition of the highest-energy cosmic-rays with energy. They claim that their analysis supports the possibility that there is a difference in mass composition between cosmic rays arriving from the Northern and Southern Hemispheres with, consequently, a diversity of sources of high-energy cosmic rays in the Northern and Southern skies. Were this deduction to be correct, it would have profound implications for theories of cosmic ray origin.

In this paper the analysis set out in SDA is examined in some detail, with additional information from other Northern and Southern Hemisphere studies also discussed. While it is found that the claim for a North/South difference is not established, the conclusion of SDA that the mean mass becomes heavier as the energy increases is convincing. This statement is itself of considerable importance. A similar verdict on the mass dependence with energy, based on some of the data published by the Telescope Array group, was reported earlier by the present author (Watson 2019). While neither analysis permits an estimate of the mean mass, the conclusions are consistent with those of the Pierre Auger Collaboration (Aab et al. 2014a and Aab et al. 2017). Comparisons of these data made against the results of simulations showed that protons gradually disappear from the cosmic ray beam above ~2 EeV, with helium and the CNO group becoming increasingly dominant at higher energies (Aab et al. 2014b).

## 2. The Elongation Rate Method and the Data Sets used

The material examined in the SDA study comes from observations made by three collaborations working at two sites in Utah, USA, and one collaboration in Malargüe, Argentina. The data comprise measurements of the mean depth at which the rate of dissipation of energy by air showers reaches a maximum, $X_{max}$. For a primary cosmic ray of 20 EeV, typical of the events discussed here, the rate of energy loss is ~30 PeV $g^{-1}$ $cm^2$, with the mean value of $X_{max}$ being ~750 g $cm^{-2}$. It is observed that the average value of $X_{max}$, $<X_{max}>$, increases with energy. It is customary to summarise such measurements using the concept of the Elongation Rate introduced by Linsley (1977), which describes the rate of change of $<X_{max}>$ with energy, measured in units of g $cm^{-2}$ per decade of energy. While to estimate the mass at a given energy, measurements of $<X_{max}>$ must be compared with predictions from hadronic models, to search only for *changes* in the mean mass it is useful to study the behaviour of Elongation Rate itself. This latter approach is a more reliable and straight-forward way to address the question: does the mean mass change with energy?

When there is only one mass-species present in the primary beam, the Elongation Rate (ER) estimated in many models is ~60 - 70 g $cm^{-2}$ per decade, with the difference in $<X_{max}>$ between showers initiated by protons or iron nuclei being ~80 – 100 g $cm^{-2}$. It follows that a change from a pure iron beam to a pure proton beam over a range of energy would be mirrored in a larger Elongation Rate while conversely, were the mean mass to become heavier, the Elongation Rate would fall. This happens because, as Linsley (1977) pointed out in his seminal paper, a key factor driving the Elongation Rate is the energy of the photons from the decay of neutral pions, and the mean value of this energy is lower in nucleus-nucleus collisions than in proton-nucleus collisions at the same incident energy.

In this paper only projects that have provided data above an energy of 1 EeV are discussed. This is because the questions to be addressed here are whether there is a change in the Elongation Rate above ~ 3 EeV and whether or not there is a difference between the Elongation Rates found in the Northern and Southern hemispheres at the highest energies. Only data available in refereed journals have been used. The measurements discussed in SDA include those from the Fly's Eye project – the first experiment from which reliable measurements of $X_{max}$ were obtained (Bird et al 1993), the HiRes studies (Abbasi et al. 2010) and those by the Telescope Array Collaboration from the Middle Drum (Abbasi et al 2015) and the Black Rock and Long Ridge locations (Abbasi el 2018). Another data set used by SDA, described as being from a TA hybrid trigger, was not cited in their paper and so is not considered here.

SDA chose to show the data, without any uncertainties, and their estimates of the Elongation Rates were obtained from averages of the $X_{max}$ values made above and below a break energy of 3 EeV, found by inspection after averaging all results in small bins of energy. Such a procedure assigns equal weight to each measurement: in the discussion below it is pointed out that this may not be appropriate.

The data from work carried out in Utah are shown in figure 1. Only those from Bird et al. above 1 EeV are presented as all other measurements from the Utah sites are above this energy. Similarly, in the discussions of other measurements set out below, lower energy measurements have been excluded. While this restriction weakens the accuracy of determinations of the Elongation Rate at lower energies as the energy range is small, it is the evaluations at the higher energies that are important in the context of the SDA claim.

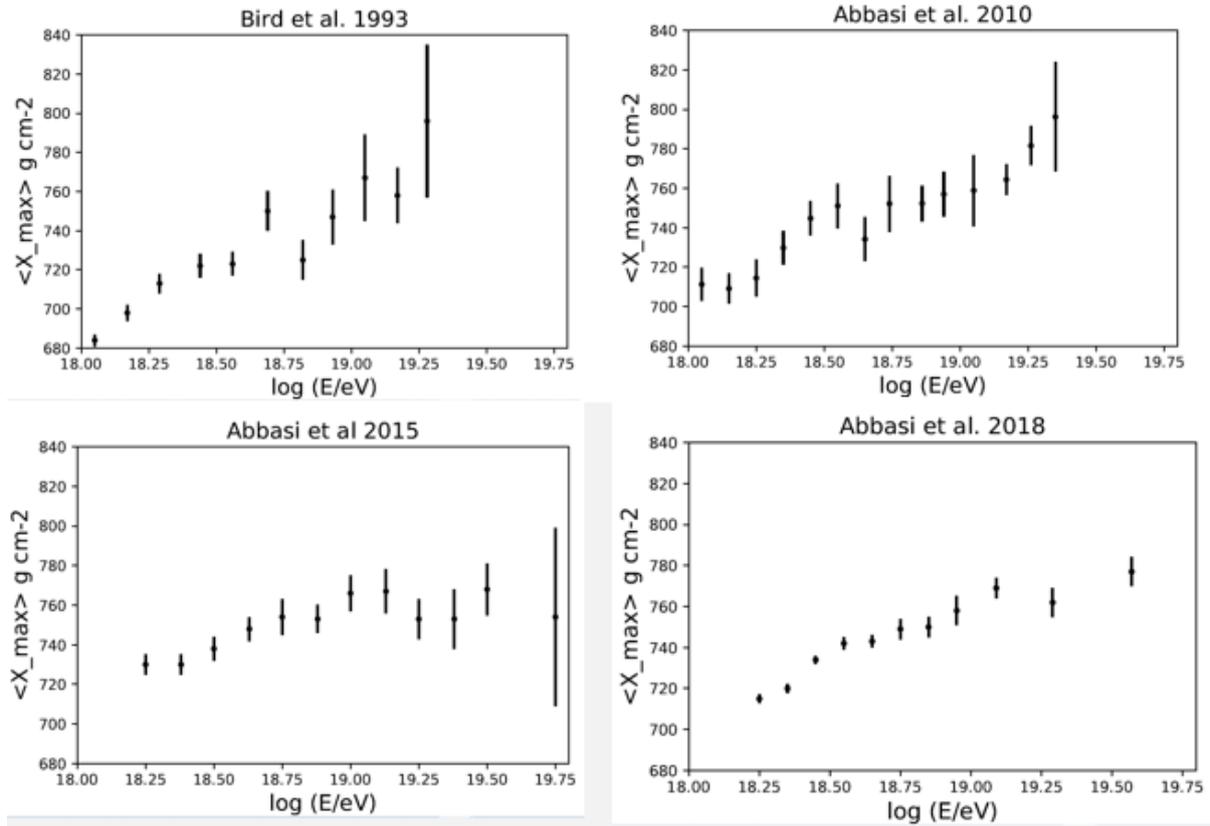

**Figure 1:** $\langle X_{max}\rangle$ as a function of energy. Data from Bird et al 1993, Abbasi et al. 2010, Abbasi et al 2015 and Abbasi et al. 2018. The energy scales are determined from the fluorescence observations.

Except in the case of the results from Black Rock and Long Ridge, where a tabulation was available, the points were measured from diagrams in the papers cited. The number of events in each set is ~1000, 814, 813 and 3330 respectively. The aggregate in the Bird et al. work is an estimate as this information is not given the paper. A break in the evolution of $X_{max}$ from a straight line is apparent, by eye, in the results of Abbasi et al. 2018, and perhaps in the Bird et al. results.

Fits have been made to search for breaks in each data set and are given in Table 1, together with estimates of the Elongation Rate across the whole energy range (ER), and for the ranges below and above the breaks fitted for each data set (ER1 and ER2 respectively).
3

|  | Number of events | $\chi^2$/ndf | ER g cm$^{-2}$/dec | $E_{break}$ log E/eV | ER1 g cm$^{-2}$/dec | ER2 g cm$^{-2}$/dec |
|---|---|---|---|---|---|---|
| Bird et al 1993 | ~ 1000 | 12.2/9 = 1.4 | 74 ± 7 | 18.6 ± 0.2 | 86 ±11 | 43 ±32 |
| Abbasi et al 2010 | 814 | 7.4/12 = 0.62 | 54 ± 6 | 18.6 ± 0.1 | 70 ± 18 | 56 ± 23 |
| Abbasi et al 2015 | 813 | 6.7/10 = 0.67 | 33 ± 6 | 18.6 ± 0.2 | 29 ± 31 | 17 ± 12 |
| Abbasi et al 2018 | 3330 | 27.4/9 = 3.04 | 54 ± 3 | 18.49 ± 0.06 | 95 ± 14 | 37 ± 5 |

**Table 1:** The source of each data set is given in column 1 with the number of events in column 2: that for Bird et al. 1993 is an estimate (see text). In column 3 the $\chi^2$ values per degrees of freedom ($\chi^2$/ndf) for fits to single straight lines across the whole energy range are shown with these values (ER) given in column 4. The break energy is listed in column 5, while the estimates of the elongation rates below and above this energy are in columns 6 and 7 (ER1 and ER2).

Several conclusions can be drawn from the summary in this table:

1. Given the different systematic uncertainties that are likely to be involved, arising from detector calibrations and atmospheric monitoring for example, the agreement between the independent measurements, with the exception of the set from Abbasi et al. 2015, is striking.
2. Except in the case of data from Abbasi et al 2018, a single Elongation Rate is an adequate description of the data in each case, although the computed values of ER differ by amounts outwith statistical uncertainties.
3. For data other than those from for Abbasi et al. 2015, the Elongation Rates below the break (ER1) are larger than those above the break (ER2), although the statistical significance is strong only in the Abbasi et al. 2018 work.
4. The values of ER1, again with the exception of that from Abbasi et al. 2015, are in agreement within the statistical uncertainties.

Conclusion 3 was reached previously from a study of Utah data (Watson 2018).

**3. Is there a North-South difference?**

The data from the fluorescence work summarised in Table 1 have been collected over a period of nearly 40 years. The agreement is generally impressive, with the exception of the Middle Drum results (Abbasi et al. 2015) which seem out of line with the other measurements. No comment has been made by the Telescope Array Collaboration about the difference between this result and that of Abbasi et al. 2018, both carried out relatively recently at Millard County, Utah, or with the two earlier reports from the Utah Dugway location (Bird et al. 1993 and Abbasi et al. 2010).

The spreads of the measurements and of the uncertainties from the different studies are evident in figure 2 where the four plots of figure 1 are put together. Unweighted averaging of these data seems inappropriate. Excluding the Middle Drum results, the weighted averages derived from the other measurements are ER1 = 93 ± 4 g cm$^{-2}$/decade and ER2 = 38 ± 5 g cm$^{-2}$/decade. The value of ER2 is smaller than that derived by SDA (48 ± 10 g cm$^{-2}$/decade) because of the high weight of the measurement with the greatest number of events (Abbasi et al. 2018).

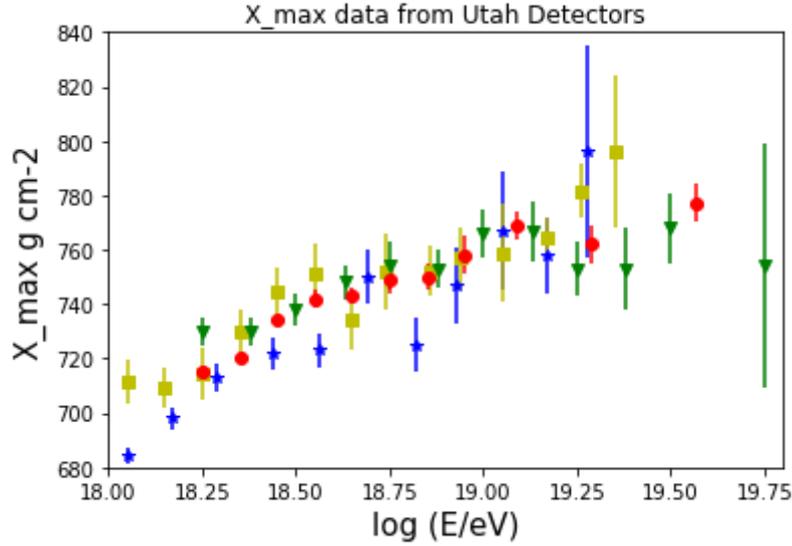

**Figure 2:** The data of figure 1 are shown together so that the spreads in the measurements, and in particular in the uncertainties, can be readily compared: 'stars': Fly's Eye, Bird et al. 1993; 'squares': HiRes, Abbasi et al. 2010; 'triangles': Middle Drum, Abbasi et al. 2014; 'circles': Black Rock and Long Ridge, Abbasi et al. 2018.

The values of ER1 and ER2 are clearly different, supporting the conclusion from the Pierre Auger Collaboration (Aab et al. 2014a) that there is a break in the Elongation Rate and that mean mass of cosmic rays increases with energy above this break, as discussed in detail in Aab et al. 2014b.

The break energies reported in the two measurements are log E/EeV = 18.27 ± 0.04 (Auger Collaboration, Aab et al. 2014a) and log E/EeV = 18.49 ± 0.06 (Telescope Array, Abbasi et al. 2018). The difference in the these energies does appear to be significant and is larger than the 10% shifts of energy scale thought to exist between estimates of energy from the two collaborations (Deligny et al 2019). However, in view of the corroborating evidence now offered by SDA for a change of mean mass with energy, the approach of the Telescope Array group to determining the energy surely needs revision as, unlike estimates of primary energies made by the Auger Collaboration (Aab et al. 2020) which are independent of prior knowledge about mass composition or about hadronic interactions, the approach of the Telescope Array Collaboration requires assumptions to be made about both of these variables. In particular it should be noted that a pure proton beam was assumed in the Telescope Array analysis, a result inconsistent with the value of ER1 found from the Utah data.

Over the 40 years of measurements made by several teams working in Utah, the understanding and precision of the technique have surely improved. For example, atmospheric monitoring, which was somewhat limited in the early days, is now more sophisticated. It therefore seems most appropriate to consider the last report (Abbasi et al. 2018) from the newest detectors at Black Rock and Long Ridge as being the most reliable. These two instruments have yielded more events than from the earlier three studies combined, with 804 being above the break energy. As well as the larger number of events, the accuracy of reconstruction of $X_{max}$ is reported by Abbasi et al. (2018) to be superior to that of Middle Drum work and to what was achieved in the earlier studies.

The data from Abbasi et al. 2018, taken from the tabulation in that paper (see also figure 1), are plotted again in figure 3.

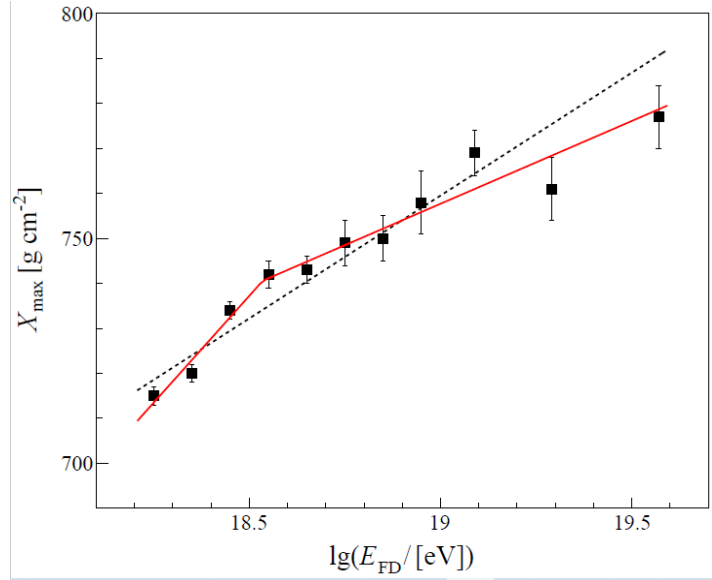

**Figure 3:** $X_{max}$ vs Energy as reported by Abbasi et al. 2018. The dashed line is a fit to the points with a single straight line (Elongation Rate = 54 ± 3 g cm$^2$/ decade) while the full line is a fit with a break and the best-fit straight lines found simultaneously (ER1 = 95 ± 14 and ER2 = 37 ± 5 g cm$^{-2}$/decade of energy).

Two lines are shown. The dashed line is a fit assuming that the data can be described by a single line. That this fit is poor is evident by eye, as is borne out by the value of $\chi^2/9 = 3.04$ that implies that the probability of a single line fit being adequate is $< 10^{-3}$. Abbasi et al. 2018 reported a $\chi^2/9 = 10.67$ which has a probability of an adequate fit ~0.3[1].

The claim of SDA that there is a difference of Elongation Rate in the Northern and Southern hemispheres rests on whether the value of ER2 derived from the Abbasi et al. 2018 work of 37 ± 5 g cm$^{-2}$/decade is different from the Auger Collaboration result of 26 ± 2.5 g cm$^{-2}$/decade (Aab et al. 2014a). The figure of 37 ± 5 g cm$^{-2}$ is lower than the average estimate reported in SDA for the reasons already stated. The differences are within ~2 sigma and therefore far from the 5 sigma result that would be required for the important claim that the mean mass of cosmic rays in the two hemispheres is different to be credible (see Table 2). There is thus no convincing evidence to support the SDA assertion.

## 4. Other measurements of the Elongation Rate from the Northern and Southern Hemispheres

Other measurements from which Elongation Rates can be derived have been reported from both the Northern and Southern Hemispheres. These support the contention that no differences between the Elongation Rates measured in the two hemispheres have yet been established.

### 4.1 Measurements by the Yakutsk group using air-Cherenkov light

The Yakutsk group has studied high-energy cosmic-rays for over 40 years and Knurenko and Petrov (2019) have recently reported a summary of their measurements of $X_{max}$. Their estimates come from studies of the lateral distribution of the air-Cherenkov radiation associated with air-showers. Such measurements, which are likely to have different systematic uncertainties from those associated with measurements of fluorescence radiation, also provide a powerful tool for determining the energy of the primary particles and have been exploited in this manner by Russian groups since the 1960s.

---

[1] It is not yet known why the value reported by Abbasi et al (2018) is so different from that quoted here, although this issue was pointed out some time ago (Watson 2019). It seems possible that the bin widths used in the energy scale on the x-axis were included in the $\chi^2$ calculation. The bin widths were 0.1 in log E up to log E(EeV) = 19.00, and 0.2 for the points at the three highest energy points. Under this assumption, M Unger has computed $\chi^2$ using a ROOT routine, finding a value of 11.4, close to that of the Telescope Array of 10.67. The 7% difference may arise because more digits were used in the original fit than are tabulated in the paper.

In a summary plot in Abbasi et al. 2018 (figure 2 of that paper) the Yakutsk measurements are seen to be in reasonable agreement with results from the Telescope Array and the Auger Observatory and therefore their inclusion here adds additional and reliable information to the discussion. The Yakutsk data, taken from the tabulation in their paper, are shown in figure 4. It is evident that a straight line is a poor fit to the measurements although the reduced $\chi^2$ is only 1.03. Analysis shows that is a break in the evolution of the Elongation Rate at an energy of log E/eV = 18.5 ± 0.1, as is evident by eye, with ER1 and ER2 being 86 ± 18 and 28 ± 7 g cm$^{-2}$/decade respectively.

The data from the Yakutsk array again confirm that the Elongation Rate above the break is smaller than that before it, and significantly below that for a single mass species thus demonstrating again that the mean mass is increasing with energy.

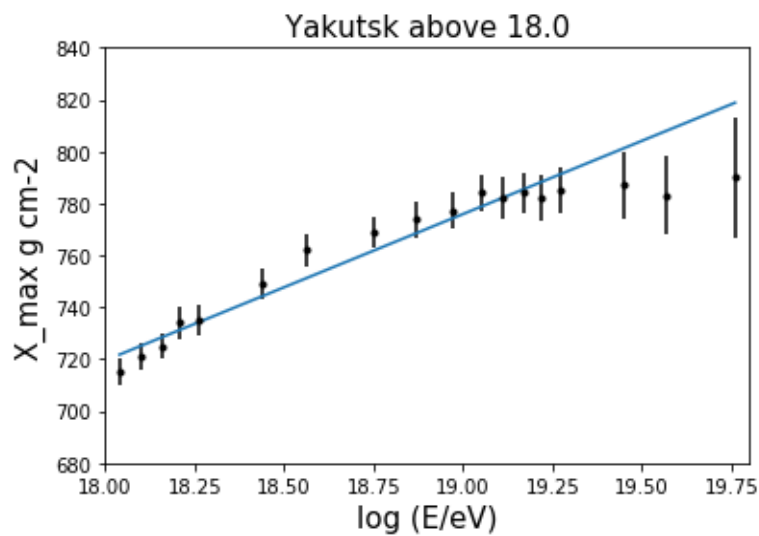

**Figure 4:** Data of $X_{max}$ vs log E/eV from Knurenko and Petrov 2019). The energy scale is derived from measurements of Cherenkov radiation produced by the particles of the air-showers.

In their paper, Knurenko and Petrov (2019) quote 50 g cm$^{-2}$/decade for ER2. This is derived (private communication) by taking the extreme limits of the estimates of the systematic uncertainties (negative at the lowest energy and positive at the highest energy) from the plot shown in figure 5. The systematic uncertainties in $X_{max}$ grow from ± 5 to ± 20 g cm$^{-2}$ over the energy range of interest. This approach would lead to a negative value for the Elongation Rate if the choice of positive and negative values was reversed. A more conventional use of the data has been adopted here.

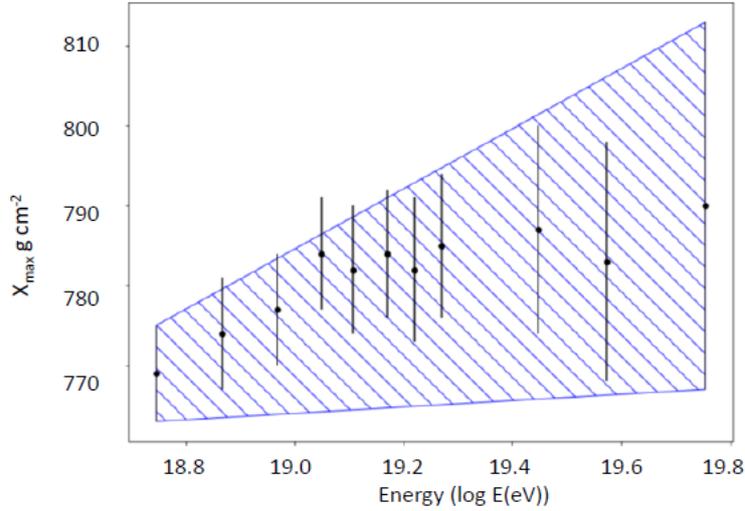

**Figure 5:** A plot of the Yakutsk data above log E/eV = 18.75, made by Knurenko and Petrov (private communication) showing the systematic uncertainties (shaded area) of their measurements which range from ± 5 to ±20 g cm$^{-2}$ over the energy range shown.

## 4.2 Measurements of the Elongation Rate made by the Auger Collaboration using the risetime of signals from water-Cherenkov detectors

It has been known for some time that measurements of the risetimes of the signals in deep, large area, water-Cherenkov detectors can give information on fluctuations in the longitudinal development of air showers (Watson and Wilson 1974). The parameter adopted to characterise the risetime has been the time taken for the signal to grow from to 10% to 50% of its final value. Using data from the Haverah Park array, it was shown, for the first time, that 'between shower' fluctuations were detectable at an energy of ~1 EeV.

The scope of this early effort has been hugely extended using the vast database accumulated by the Auger Collaboration with detectors of an identical depth to those used at Haverah Park (Aab et al. 2017). Measurements of the risetime of signals were made in over 81,000 events with energies above 0.3 EeV of which 55501 and 517 were above 1 EeV and 30 EeV respectively. Accruing so large a database was possible because water-Cherenkov detectors, unlike optical detectors, can be operated for 24 hours per day. However the risetime measurements do not give the depth of shower maximum directly and must be calibrated using those events for which information from the fluorescence detectors is also available. The calibration was made using 1.4% of the data. Details are given in Aab et al. 2017.

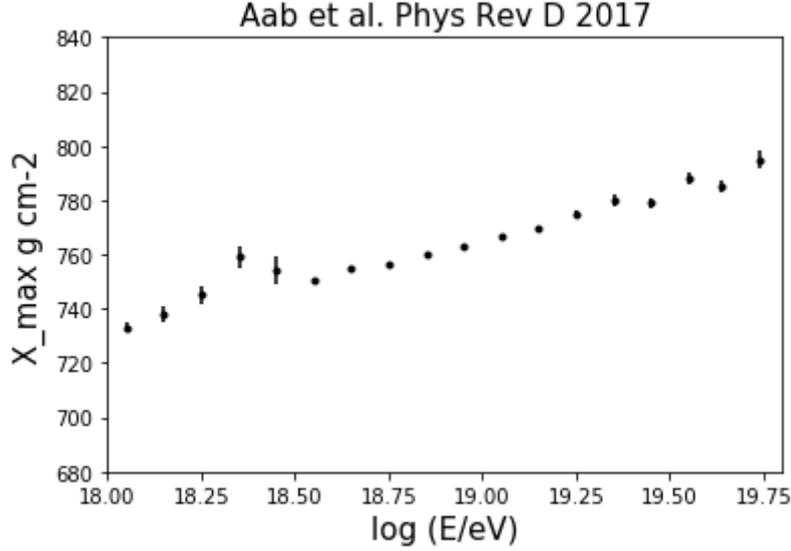

**Figure 6:** Estimates of $X_{max}$ as a function of energy made using the risetime technique (Aab et al. 2017). Only data from the 55501 events with energies above 1 EeV are shown.

The estimates of $<X_{max}>$ obtained with this method above 1 EeV are shown in figure 6. The events of energy below 3 EeV are obtained from a portion of the 3000 km² array, with an area of 24 km², where the detector spacing is 750 m rather than 1500 m. The Elongation Rate above the break energy of 3 EeV is $33.2 \pm 0.7$ g cm$^{-2}$/decade.

## 6. Discussion and Conclusions

Measurements of Elongation Rates made above 1 EeV in the two hemispheres are summarised in table 2. The statistical uncertainties are small in the cases from the Auger Observatory measurements because of the greater exposure. Such large databases also lead to a better understanding of systematic uncertainties as more tests can be undertaken.

|   | Number of events | $E_{break}$ log E/EeV | ER1 g cm$^{-2}$/dec | ER2 g cm$^{-2}$/dec | $\chi^2$/ndf |
|---|---|---|---|---|---|
| Abbasi et al 2018 | 3330 | $18.49 \pm 0.06$ | $95 \pm 14$ | $37 \pm 5$ | 7.84/7 |
| Knurenko and Petrov 2019 | 1156 | $18.50 \pm 0.10$ | $86 \pm 18$ | $28 \pm 7$ | 2.1/14 |
| Aab et al. 2014a | 12608 | $18.27 \pm 0.04$ | $98 \pm 12$ | $26 \pm 2.5$ (sys +7/-1.9) | 5.89/12 |
| Aab et al. 2017 | 55501 | $18.50 \pm 0.10$ | $66 \pm 10$ | $33.2 \pm 0/7$ | 59/15 |

**Table 2:** A summary of measurements from the Northern (NH) and Southern (SH) Hemisphere observatories using measurements above 1 EeV. Note that the two measurements from the Auger Observatory in the Southern Hemisphere are independent and were made using very different techniques. For consistency only data above 1 EeV have been used for all data sets so that, for ER1, the number of points used to define each slope is smaller than for the determinations of ER2, and the uncertainty consequently larger. The entries for $\chi^2$/ndf have been computed for the overall two-line fits. The value for the Yakutsk data is rather low, perhaps reflecting an over-estimate of the uncertainties given for the data shown in figures 4 and 5. The large value found for the Aab et al. 2017 work reflects that, with such a large sample, subtleties in the evolution of the Elongation Rate are starting to appear that show that single line fits over large energy ranges do not describe the data

adequately. This point will be discussed in a forthcoming paper in which an even larger set of risetime data will be discussed (Pierre Auger Collaboration, in preparation).

From the data shown in table 2 one can conclude:

1. The results derived from the newest and highest statistics information published by the Telescope Array (Abbasi et al. 2018) are entirely consistent with data from the three other measurements.
2. The values of ER1 are larger than those predicted using models of hadronic interactions, indicating that the mean mass is falling as the energy increases up to a break energy of ~3 EeV.
3. The fact that ER2 is smaller than ER1, implying that above the break energy the mean mass is increasing with energy.
4. It is clear that the hypothesis proposed by Sokolsky and D'Avignon (2021) of an Elongation Rate difference in the two hemispheres is not supported. The weighted averages of ER2 obtained for data from the Northern and Southern Hemispheres respectively are 34 ±4 and 32.7 ± 0.7 g cm$^{-2}$ per decade of energy. Thus the claim made in SDA of a mass difference between the cosmic rays from the two hemispheres is not substantiated.

What is clearly established is that the Elongation Rates above and below the break energy are significantly different, confirming earlier claims by the Auger Collaboration (Aab et al. 2014a). Thus, unless there is a dramatic change in features of hadronic interactions in collisions above ~3 EeV, one is driven to conclude that the mean mass of high-energy cosmic-rays becomes larger as the energy increases.

The confirmation of a change in mean mass with energy surely forces a revaluation of energy estimates made by the Telescope Array group as these estimates are mass-dependent, with protons having been assumed at all energies. It may also be an opportune time to reconsider the hadronic model to be adopted for this calculation. Such work may help an understanding of the fact that the fluxes of cosmic rays presently observed by the Auger and Telescope Array teams from the same part of the sky are in significant disagreement. From a joint study it was reported that an adjustment of ± 10% per decade in the energy scales of the two instruments is needed to obtain agreement between the fluxes above the energy where the suppression of the spectrum commences (~20 EeV). Such a shift is greater than the accuracy claimed for the individual measurements (Deligny et al. 2019). This issue is of primary importance and must be resolved before North/South differences of any kind can be discussed.

The conclusion that the particle beam becomes heavier at the highest energies is also important for predictions of the exposure required to study very high-energy neutrinos, as the presence of nuclei with masses heavier than protons reduces the neutrino flux at high energies below that predicted if the primaries are dominantly protons (e.g. Ave et al. 2005, Hooper, Taylor and Sarkar 2005).

This research did not receive any specific grant from funding agencies in the public, commercial, or not-for-profit sectors.

Declaration of competing interest
The author declares that he has no known competing financial interest or personal relationships that could have appeared to influence the work reported in this paper.

Acknowledgements
I am grateful to M. Unger and A. Yushkov for helpful discussions, and to S. Knurenko and I. Petrov for the provision of figure 5 and for clarifications of aspects of their results.